# Monte Carlo simulation of spin relaxation in nanowires and 2-D channels of II-VI semiconductors

Ashutosh Sharma *, Swetali Nimje and Bahniman Ghosh

*Department of Electrical Engineering, Indian Institute of Technology, Kanpur 208016, India*

## ABSTRACT

We have analysed spin relaxation behaviour of various II-VI semiconductors for nanowire structure and 2-D channel by simulating spin polarized transport through a semi-classical approach. Monte Carlo simulation method has been applied to simulate our model. D'yakanov-Perel mechanism and Elliot-Yafet mechanism are dominant for spin relaxation in II-VI semiconductors. Variation in spin relaxation length with external field has been analysed and comparison is drawn between nanowire and 2-D channels. Spin relaxation lengths of various II-VI semiconductors are compared at an external field of 1kV/cm to understand the predominant factors affecting spin de-phasing in them. Among the many results obtained, most noticeable one is that spin relaxation length in nanowires is many times greater than that in 2-D channel.

## INTRODUCTION

Various recent advances in the field of spintronics have opened up paths for extensive study in spin transport characteristics of various semiconductors [1-3]. Many of them are targeted to analyse the usefulness of spin transport to design a new set of devices which will be faster, smaller and more power efficient than any of the conventional semiconductor devices [4-6]. It is judged on the basis of efficiency of transferring, storing or reading the data and field of spintronics holds a lot of promise for this purpose [6-8].Efficient data transfer in a certain spintronics based device is dependent on duration for which the polarity of injected spin is retained and the distance over which this polarity is maintained which is termed as spin relaxation length. The analysis of those parameters which enable us to manipulate spin relaxation length is critical in this regard [9-11].

We are concentrating on finding out spin de-phasing lengths for nanowires and 2-D structures composed of II-VI semiconductors. The reason for choosing this class of materials is the recent growing interest in them which arises due to their remarkable electromagnetic properties [12-16]. They possess a desirable range of values of opto-electronic parameters like band-gap, mobility etc. which enables us to choose a right combination of parameters to suit our purpose. They have also shown interesting results under doping with magnetic substances, which in itself is a wide area of research that can be explored when their spin transport properties are well understood [17-19].

We study the effect of varying the electric field across 2-D channel and nanowire structures on spin relaxation length. Analysis based on these results will help us to establish an appropriate set of conditions needed for developing spintronics based devices. Fabrication techniques for the nanowires composed of II-VI semiconductors are also under development. Several significant advances have been made in comparing the advantages of various fabrication techniques in terms of nanowire properties [19-21].

Monte Carlo methods have been used by many researchers for analysing spin transport [22, 23]. In this work too, we have based our simulations on semi classical Monte-Carlo approach for modelling spin de-phasing phenomena in II-VI semiconductor nanowires and 2-D structures. This approach enables us to analyse in an orderly and step-by-step manner, the spin-orbital interactions and scattering mechanisms which occur while de-phasing and thus gives us a better insight than drift-diffusion or hydrodynamic transport model.

## MODEL

We take a semi-classical approach with multi sub-bands for analysis of spin relaxation during spin transport[24, 25, 26] and employ the Monte Carlo simulation method [22,23,24] to meet this requirement. Monte Carlo method provides an elaborate and effective approach for modelling spin transport. For the nanowire, motion along y and z axis is constrained and is only along x direction, while for the 2-D channel, it is constrained along z axis. Current is along x-direction and electrons are injected from one end with a specific spin orientation which are driven to other end under the influence of field produced by the potential difference created across the two ends. In addition to this a transverse field is also applied.

*e-mail : ashu.iitk@gmail.com, ashushar@iitk.ac.in

Due to transverse electric field, Rashba spin-orbit coupling [27] becomes effective which occurs as a result of Structural inversion asymmetry, since this field disrupts the Bulk inversion asymmetry. Bulk inversion asymmetry is responsible for Dresselhaus spin-orbit interaction [28]. Effect of these interactions on the spin of electron is according to the Hamiltonian expressions below:

(i) For nanowire structures

$$H_R^{1D} = -\eta k_x \sigma_y$$
$$H_D^{1D} = -\beta(<k_y>^2 - <k_z>^2)k_x \sigma_x$$

The energy levels of subbands in a nanowire is given by

$$\psi_{p,q}(y,z) = \sqrt{\frac{4}{L_y L_z}} \sin\left(\frac{p\pi y}{L_y}\right) \sin\left(\frac{q\pi z}{L_z}\right)$$

(ii) For 2-D channel

$$H_R^{2D} = -\eta(k_y \sigma_x - k_x \sigma_y)$$
$$H_D^{2D} = -\beta <k_z>^2 (k_y \sigma_y - k_x \sigma_x)$$

value of constant β depends on material. Value of η depends on transverse electric field and certain material properties like spin orbit splitting of valence band, electronic charge, effective mass and band gap. Spin orbit Hamiltonian can also be expressed as

$$H_B = H_D + H_R = \frac{g\mu_B}{2}\vec{\sigma}.\vec{B}$$

Using the relation $\vec{\Omega} = \frac{ge\vec{B}}{2m}$ for the precession vector $\vec{\Omega}$ in Larmor precession equation which governs spin modification during free flight of electron [26,29]

$$\frac{d\vec{S}}{dt} = \vec{\Omega} \times \vec{S}$$

We get

$$\frac{d\vec{S}}{dt} = \frac{g\mu_B}{\hbar}(\vec{B} \times \vec{S})$$

on substituting expression for components of $\vec{B}$ and $\vec{S}$ using above equations gives us following relations:

$$\frac{dS_x}{dt} = \frac{2}{\hbar}(\beta_{eff} k_y - \eta k_x)S_z$$
$$\frac{dS_y}{dt} = \frac{2}{\hbar}(\eta k_y - \beta_{eff} k_x)S_z$$
$$\frac{dS_z}{dt} = \frac{2}{\hbar}(\eta k_y - \beta_{eff} k_x)S_y - \frac{2}{\hbar}(\beta_{eff} k_y - \eta k_x)S_z$$

In D'yakanov-Perel [30] mechanism, after one iteration momentum and spin state variables of each electron are updated in accordance with the scattering mechanism during the free flight and therefore we get a distribution of states for both spin and momentum, from which we evaluate ensemble de-phasing. Elliot-Yafet mechanism causes sudden spin flip [31, 32] which is included in scattering effects as spin flip scattering [30, 33].

$$\frac{1}{\tau_s^{EY}} = A\left(\frac{k_B T}{E_g}\right)^2 \eta^2 \left(\frac{1-\eta/2}{1-\eta/3}\right)^2 \frac{1}{\tau_p}$$

$$\eta = \Delta/(E_g + \Delta)$$

$E_g$ is the band gap, $\Delta$ is spin orbit splitting of the valence band. $\tau_p$ is the total momentum relaxation time.

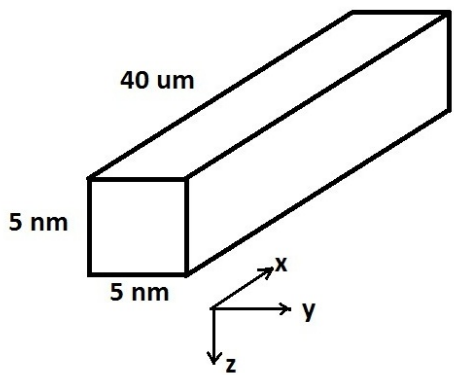

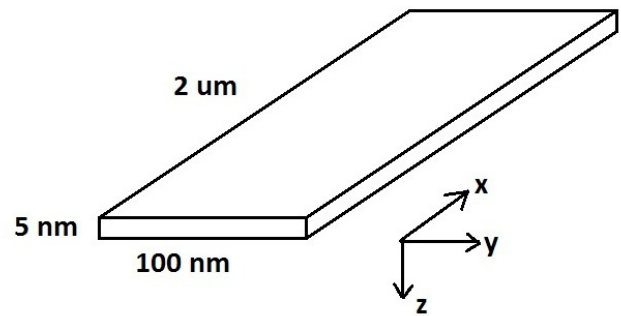

**Figure 1(a) Geometry of Nanowire**          **Figure 1(b) Geometry of 2 D Channel**

This mechanism is more dominant in materials in which value of spin orbital splitting is more. The above expression also indicates that for a material with higher band gap, momentum relaxation time will be higher resulting in comparatively more spin relaxation length, whereas higher value of spin orbital splitting will result in lesser spin relaxation length.

As shown in figure 1(a), dimensions for the nanowire are $L_X$ = 40μm, $L_y$ = $L_z$ = 5nm. Figure 1(b) shows the structure of 2-D channel whose dimensions are $L_X$ = 2μm, $L_y$ = 100nm, $L_z$ = 5nm. Doping density is 1 x $10^{25}$ /m$^3$. The magnitude of transverse electric field is 100kV/cm [16]. Simulations are run for 800,000 iterations in order to let electrons reach a steady state for consistency of results and the data is recorded and calculations are done for the final 40,000 iterations. In each iteration, spin components are updated for small time duration 'dt', which is 0.2fs in our simulations.

Spin relaxation length is the distance from the source (x=0) to the point where the ensemble spin average magnitude reduces to 1/e times of its initial value at the time of injection. The electrons are injected with an initial polarization value=1 along z-axis at x=0. Ensemble average of spin vector is calculated component-wise for the recorded data after reaching steady state using the following expression [25]

$$<S_i>(x,t) = \frac{\sum_{t=t1}^{t=T} \sum_{n=1}^{n_x(x,t)} s_{n,i}(t)}{\sum_{t=t1}^{t=T} n_x(x,t)}$$

i = x, y or z, $n_x(x,t)$ is the total number of electrons at position x, within distance Δx at time t, $s_{n,i}(t)$ represents the value of the spin component of n$^{th}$ electron at time "t". Here "T" is the end time and "t1" is the time at which we start recording the data. The magnitude of the average spin vector is given by following expression

$$|<S>(x,T)| = \sqrt{<S_x>^2 + <S_y>^2 + <S_z>^2}$$

Of the various scattering mechanisms, we have incorporated those mechanisms which make our model most appropriate for understanding spin relaxation in II-VI semiconductors. These are acoustic phonon scattering [34,35], polar optical phonon scattering [25, 35, 36], surface roughness scattering [25, 34,37] and spin flip scattering [25, 33].

Following table lists the parameters used for our simulations for various II-VI semiconductor materials [38-44].

**Table 1: Parameters Used For Simulation**

| Parameter | ZnO | ZnS | ZnSe | ZnTe | CdS | CdSe | CdTe |
|---|---|---|---|---|---|---|---|
| Band gap (in eV at 300K) | 3.37 | 3.63 | 2.7 | 2.25 | 2.42 | 1.74 | 1.49 |
| Density(g/cm$^3$) | 5.67 | 4.08 | 5.42 | 5.72 | 4.82 | 5.81 | 5.75 |
| Speed of sound (cm/s) | 6.03x10$^5$ | 5.87x10$^5$ | 4.58x10$^5$ | 3.9x10$^5$ | 4.29x10$^5$ | 3.56x10$^5$ | 3.37x10$^5$ |
| Static dielectric constant ($\varepsilon_s$) | 8.2 | 8.32 | 9.2 | 9.7 | 8.9 | 9.56 | 10.5 |
| Density of states effective mass (in m$_o$) | 0.17 | 0.13 | 0.14 | 0.154 | 0.165 | 0.13 | 0.09 |
| Spin Orbit Splitting(eV) | -0.019 | 0.092 | 0.45 | 0.93 | 0.065 | 0.42 | 0.92 |
| Acoustic phonon deformation potential (eV) | 3.83 | 4.9 | 4.5 | 7.76 | 14.5 | 11.5 | 9.5 |
| Polar optical phonon energy (eV) | 0.05 | 0.043 | 0.03 | 0.026 | 0.038 | 0.026 | 0.021 |
| Non-parabolicity factor(eV$^{-1}$) | 0.66 | 0.69 | 0.67 | 0.41 | 0.53 | 0.128 | 0.3 |
| Lande g-factor | 1.96 | 1.885 | 1.15 | -0.4 | 1.75 | 0.54 | -1.59 |

# RESULTS

**Table2: Spin relaxation lengths at 1kV/cm at 300K for various II-VI semiconductors**

| Material | Nanowire (in µm) | 2-D Channel (in µm) |
|---|---|---|
| ZnO | 26.88 | 1.632 |
| ZnS | 29.6 | 1.672 |
| ZnSe | 25.6 | 1.648 |
| ZnTe | 16.96 | 1.232 |
| CdS | 25.68 | 1.472 |
| CdSe | 22.64 | 1.28 |
| CdTe | 12.32 | 0.896 |

Figures 4-10 show us the curves followed by ensemble average spin magnitude along the length of channel at an external driving field of 1kV/cm at 300 K. Figures 4(a) - 10(a) show the decrement of ensemble spin magnitude for nanowire structures for various materials whereas figures 4(b) - 10(b) show us its reduction pattern for 2-D channel for different materials. The injected electrons have spin orientation towards z-axis and as the electrons move farther along the channel, their spin dephases due to the aforementioned spin relaxation mechanisms and the ensemble spin magnitude follows the curves as shown in figures 4-10.

These results have been tabulated in Table 2 and plotted in bar graphs in figures 2 and 3. Observing these we can easily say that the spin relaxation length for nanowire structure is significantly greater than that for 2-D channel for all the materials. This difference arises due to difference in spin relaxation mechanism. D'yakanov-Perel mechanism gets suppressed in nanowire structure [25, 45, 46] whereas in a 2-Dchannel it occurs normally. So its contribution in spin relaxation is very low in nanowire compared to 2-D channel due to which nanowire has larger spin relaxation length than a 2-D channel. Since for a 2-D channel, dimensions are larger allowing more free space, random motion of an electron increases. As a result, its rate of collision also increases resulting in more abrupt and larger changes in momentum. Now due to the spin orbit coupling effect, change in spin polarization also increases resulting in faster spin relaxation. Compared to nanowire which is one dimensional, a 2-D channel offers more freedom of motion to electron and so its motion is more random in 2-D channel. Hence by the above reasoning, spin relaxation is more and faster for 2-D channel.

From figures2and3, we also observe that for ZnTe and CdTe, spin relaxation length is very less than other compounds. This can be attributed to a significant difference in their material properties compared to other compounds. From Table 1 we can easily observe that value of spin orbital splitting for them is 0.93 and 0.92 respectively which is significantly higher than other compounds. As a result Elliot-Yafet spin relaxation mechanism becomes dominant in these compounds, which leads to faster spin relaxation. Also the value of band gap is lowest for ZnTe in Zn group and for CdTe in Cd group, which contributes to a lower value of spin relaxation length for them.

For the remaining 5 compounds, values of spin relaxation length are very similar however it is slightly lesser for CdSe. This is mainly because of a comparatively lower value of band gap and non-parabolicity factor and a slightly higher value of spin orbital splitting. The value of spin relaxation length for CdSe is not as low as CdTe and ZnTe since the value of spin orbital splitting for CdSe is much less compared to ZnTe and CdTe. Lower value of non-parabolicity factor implies lower value of momentum according to the following expression:

$$\varepsilon(1 + \alpha\varepsilon) = \frac{\hbar^2 k^2}{2m}$$

$\varepsilon$ is the electron energy and $\alpha$ is the non-parabolicity.[22, 23]. This subsequently results in a lower drift velocity and hence lesser value of spin relaxation length.

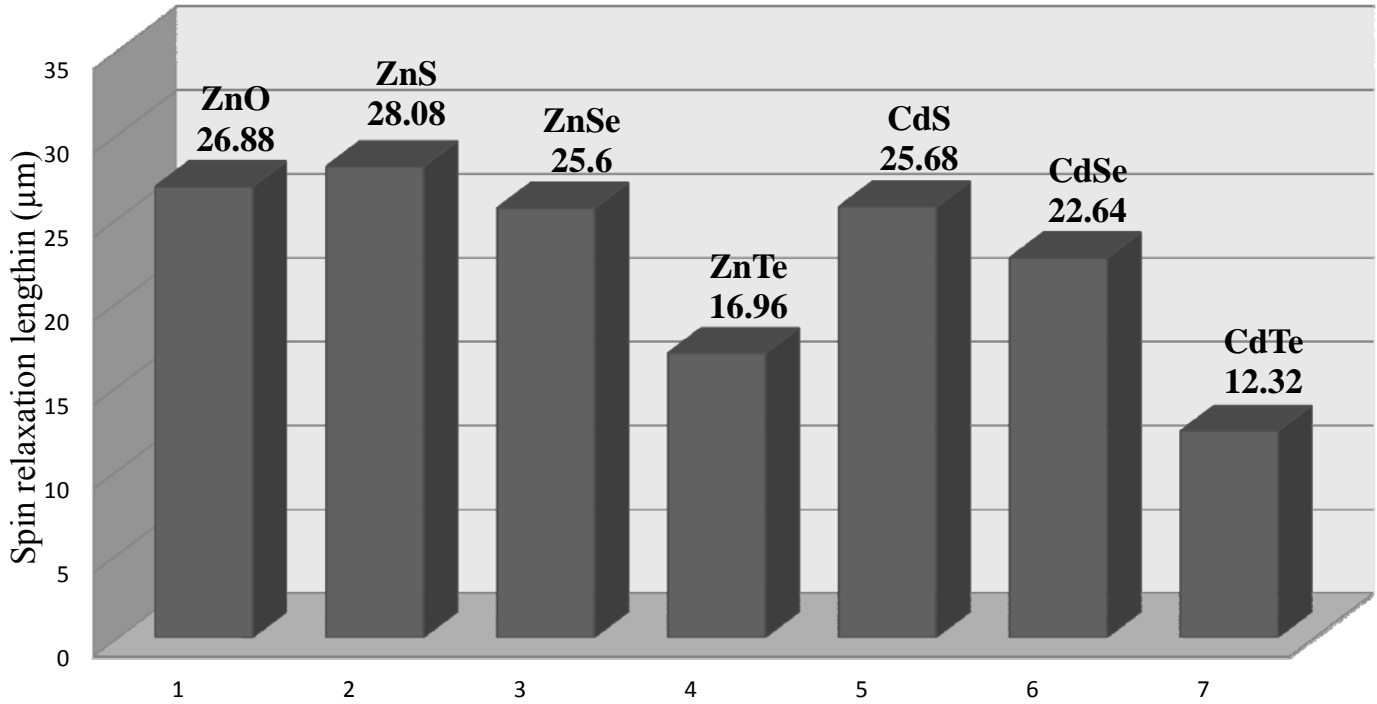

Fig.2. Bar graph showing spin relaxation length (in µm) at 1kV/cm at 300 k for nanowire

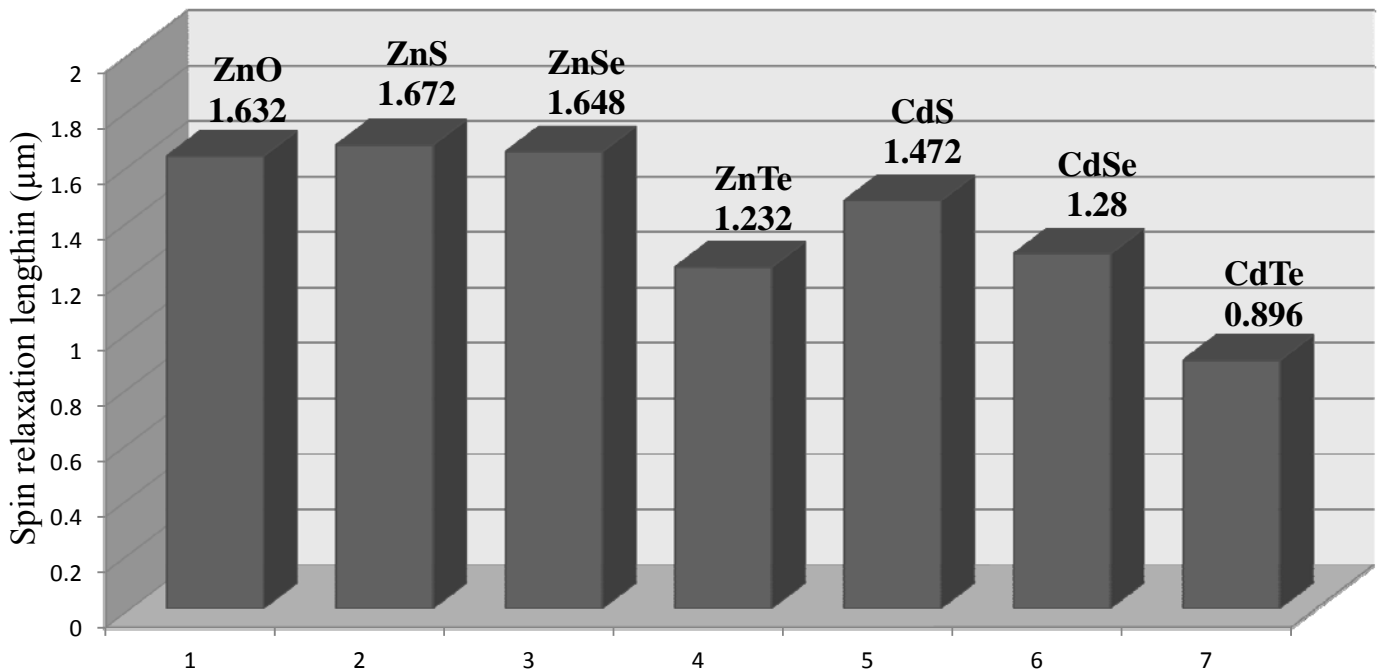

Fig.3. Bar graph showing spin relaxation length (in µm) at 1kV/cm at 300 k for 2-D channel

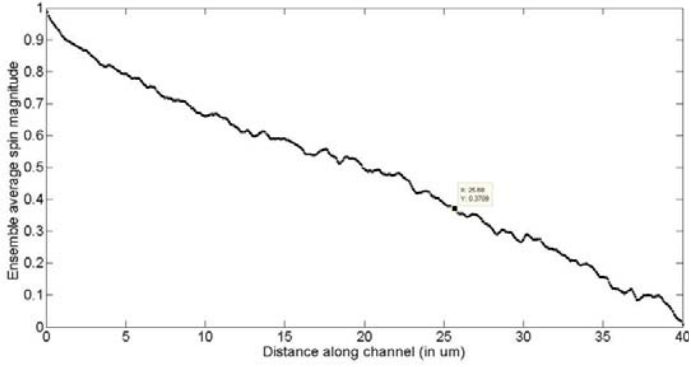 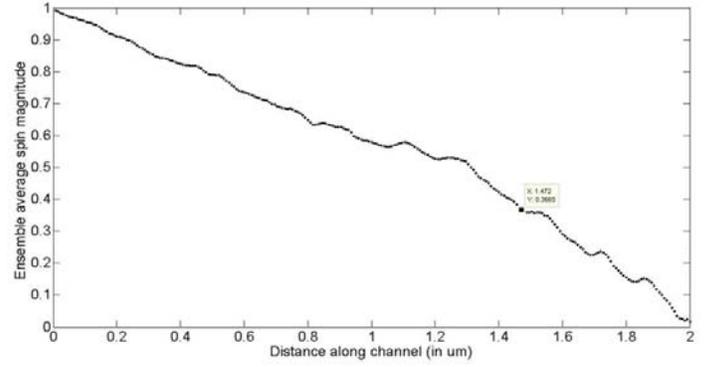

(a) (b)

Fig.4. Ensemble average spin magnitude vs. length along channel for CdS at external field of 1kV/cm at 300K for (a) nanowire and (b) 2-D channel structures

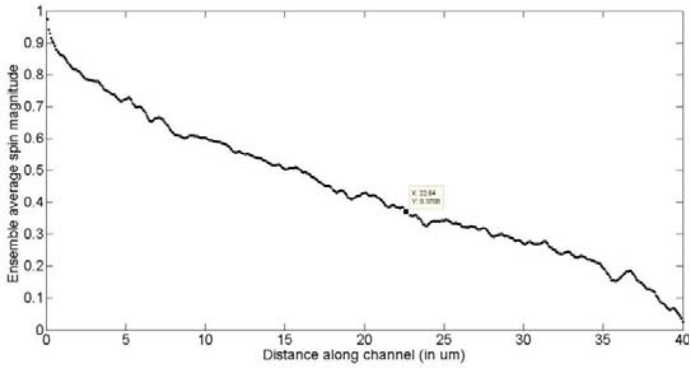 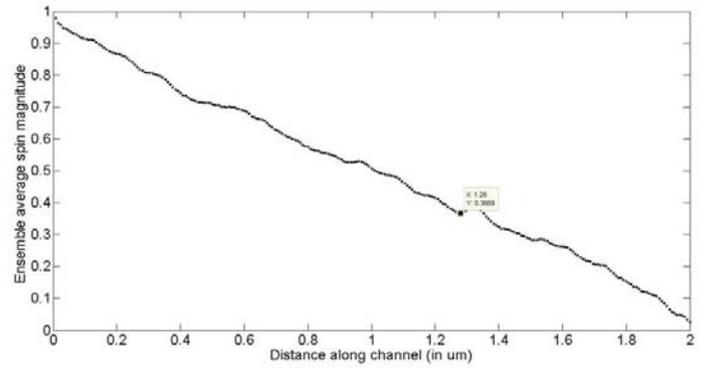

(a) (b)

Fig.5. Ensemble average spin magnitude vs. length along channel for CdSe at external field of 1kV/cm at 300K for (a) nanowire and (b) 2-D channel structures

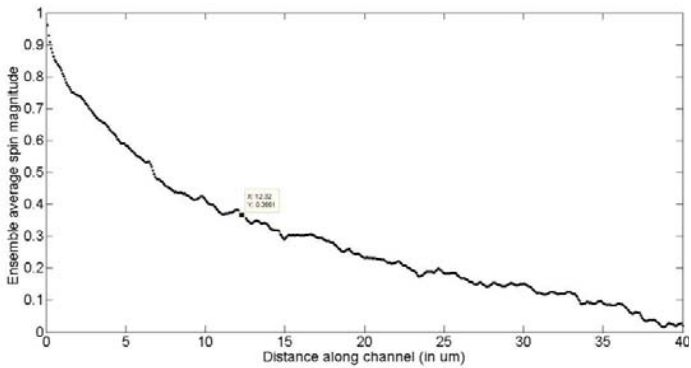 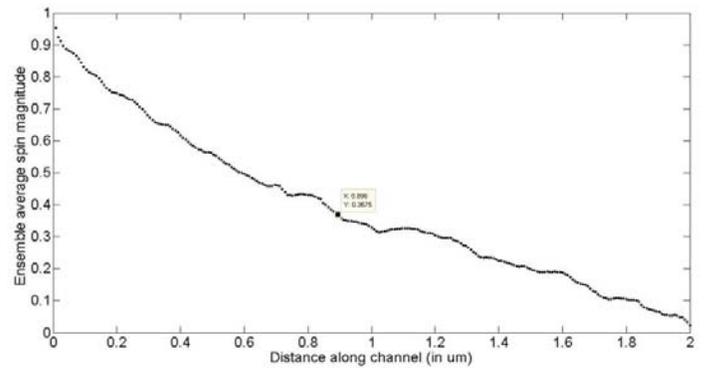

(a) (b)

Fig.6. Ensemble average spin magnitude vs. length along channel for CdTe at external field of 1kV/cm at 300K for (a) nanowire and (b) 2-D channel structures

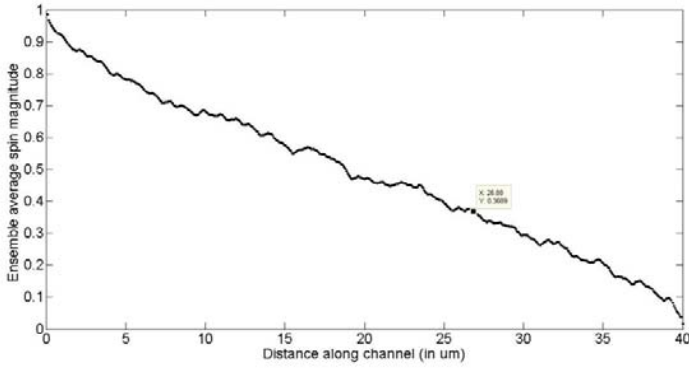 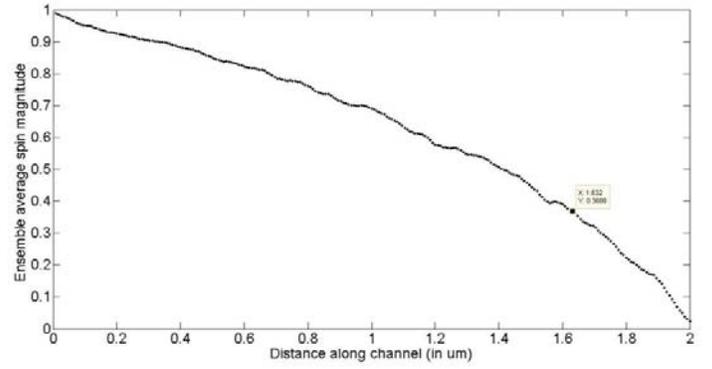

(a) (b)

Fig.7. Ensemble average spin magnitude vs. length along channel for ZnO at external field of 1kV/cm at 300K for (a) nanowire and (b) 2-D channel structures

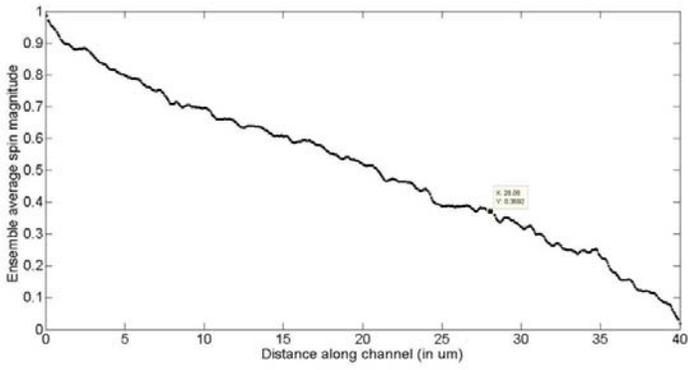 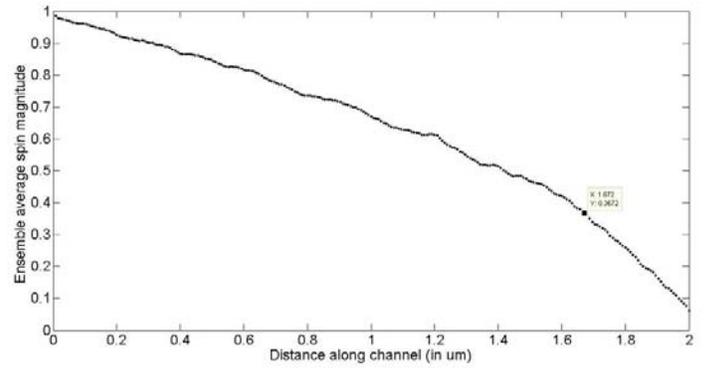

(a) (b)

Fig.8. Ensemble average spin magnitude vs. length along channel for ZnS at external field of 1kV/cm at 300K for (a) nanowire and (b) 2-D channel structures

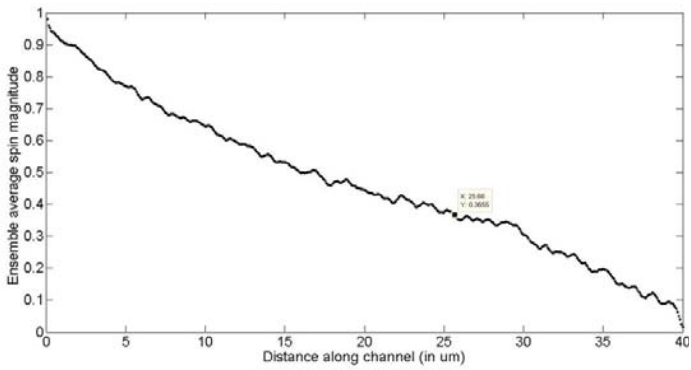 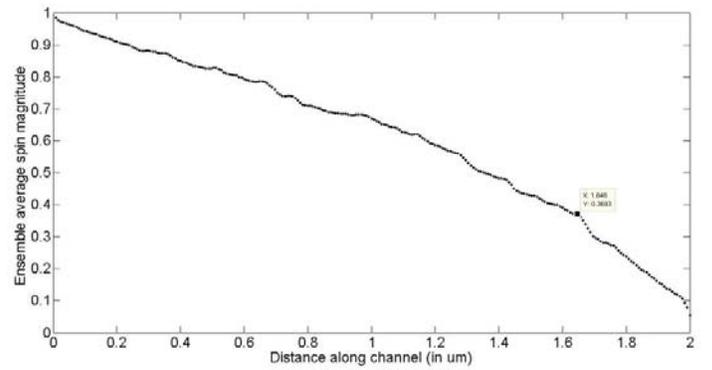

(a) (b)

Fig.9. Ensemble average spin magnitude vs. length along channel for ZnSe at external field of 1kV/cm at 300K for (a) nanowire and (b) 2-D channel structures

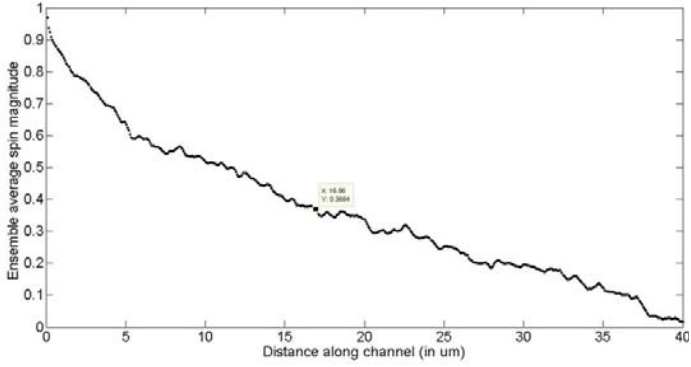 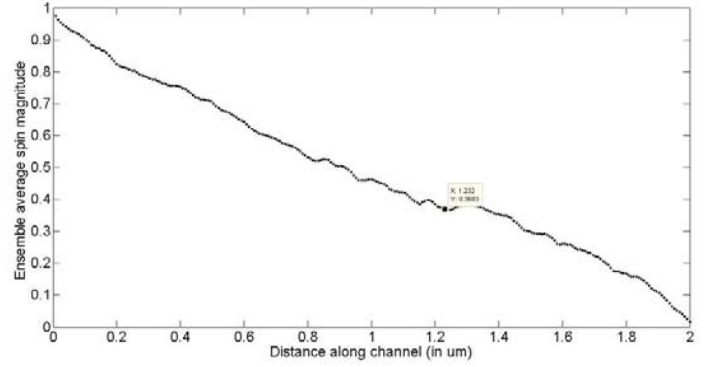

(a)                                                     (b)

Fig.10. Ensemble average spin magnitude vs. length along channel for ZnTe at external field of 1kV/cm at 300K for (a) nanowire and (b) 2-D channel structures

Variation of relaxation field with externally applied driving field is shown in figures 11-17. Relaxation length keeps on rising steadily in 2-D channel as we increase the field irrespective of the semiconductor material. This behaviour is attributed to consistent rise in drift velocity as we increase field. However the scattering rates do not continue to increase in 2-D channel with rising field and tend to reach their saturation values quickly [25].

In nanowires, the variation of spin relaxation length is not monotonic. For very low values of field it rises, since the drift velocity rises steeply in that range. At mid-range of field, although drift velocity continues to rise, its rate of increase becomes lesser resulting in a maxima followed by a minima. At high values of electric field (~10 kV/cm) Cd compounds show a drop in relaxation length whereas Zn compounds show a rise in relaxation length. This happens because acoustic phonon scattering rates are higher for Cd compounds compared to Zn compounds, which directly depend on value of acoustic phonon deformation potential which is sufficiently larger for Cd compounds than Zn compounds (refer to Table 1) and this difference becomes dominant in the high electric field range. So such increments in scattering rates dominate effect of drift velocity accordingly. This is clearly indicated by the following expression:

$$\Gamma_{nm}^{ac}(k_x) = \frac{\Xi_{ac}^2 k_b T \sqrt{2m^*}}{\hbar^2 \rho v^2} D_{nm} \frac{(1 + 2\alpha\varepsilon_f)}{\sqrt{\varepsilon_f(1 + \alpha\varepsilon_f)}} \Theta(\varepsilon_f)$$

Where $\Xi_{ac}$= acoustic deformation potential, $\rho$= crystal density, $v$= sound velocity and $\Theta$= Heaviside step-function. $D_{nm}$ is the overlap integral associated with the electron-phonon interaction and is given by [34]

$$D_{nm} = \iint |\psi_n(y,z)|^2 |\psi_m(y,z)|^2 dy dz$$

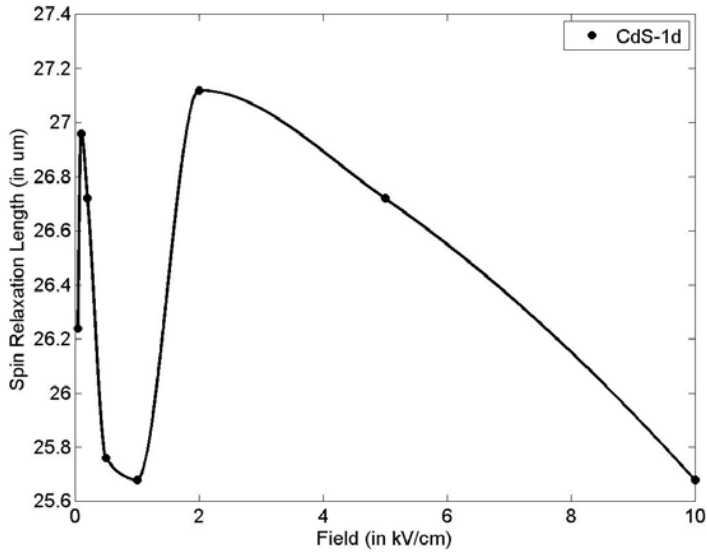 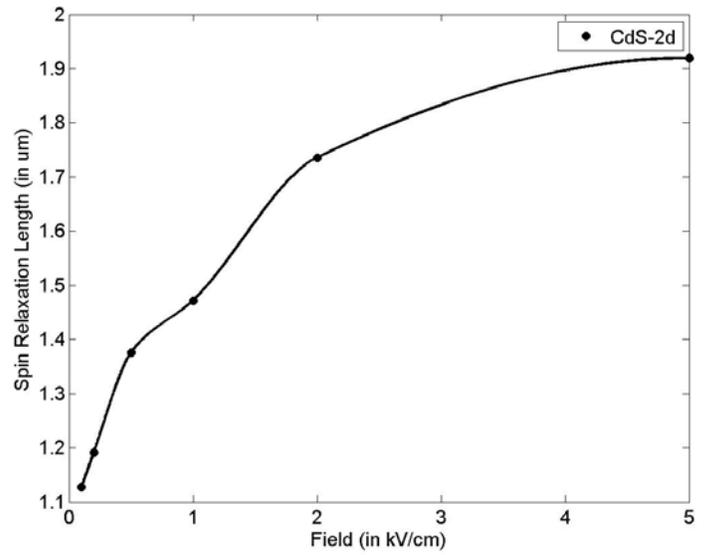

(a)                                                      (b)

Fig.11. Variation of spin relaxation length with driving electric field for CdS (a) 1-D nanowire (b) 2-D channel for injected electrons polarized along the z-direction at T = 300K

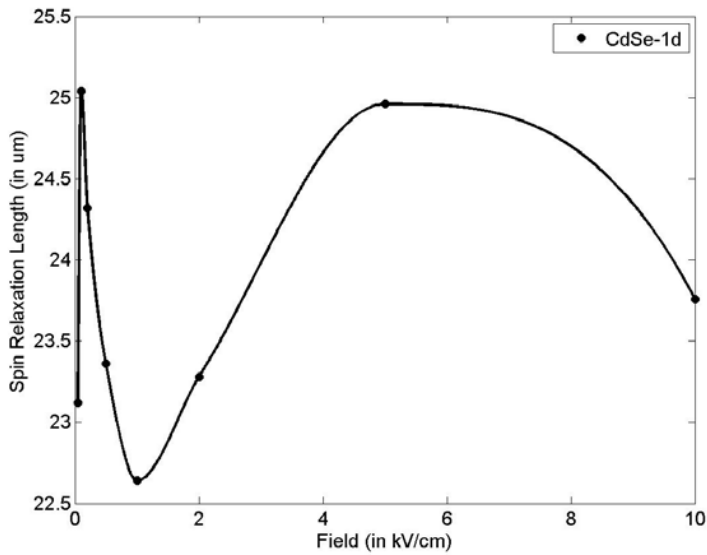 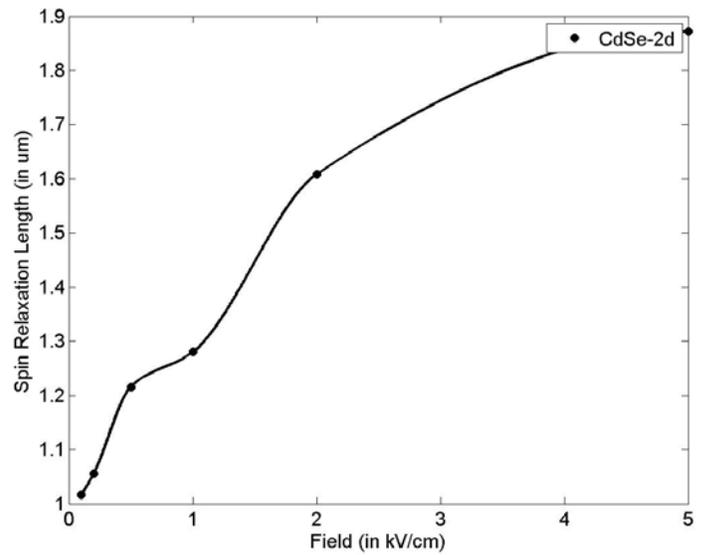

(a)                                                      (b)

Fig.12. Variation of Spin relaxation length with driving electric field for CdSe (a) 1-D nanowire (b) 2-D channel for injected electrons polarized along the z-direction at T = 300K

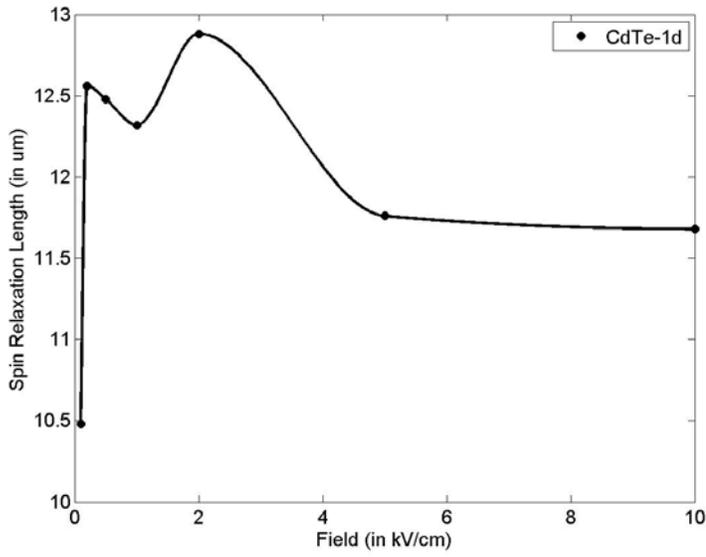 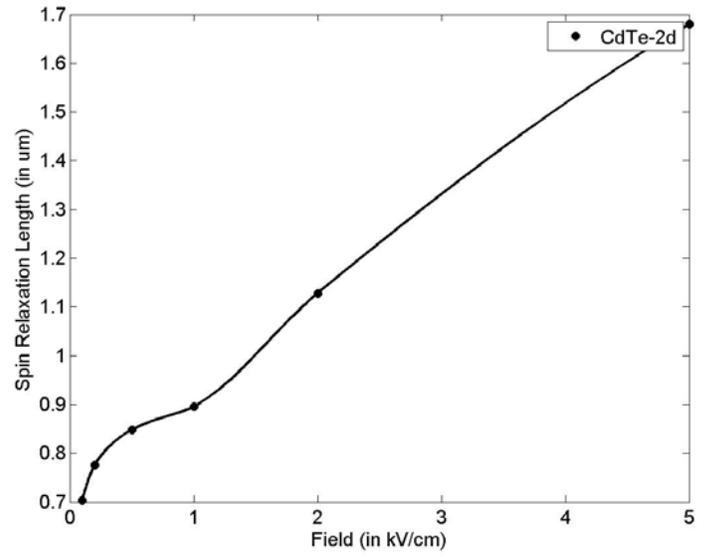

(a) (b)

Fig.13. Variation of Spin relaxation length with driving electric field for CdTe (a) 1-D nanowire (b) 2-D channel for injected electrons polarized along the z-direction at T = 300K

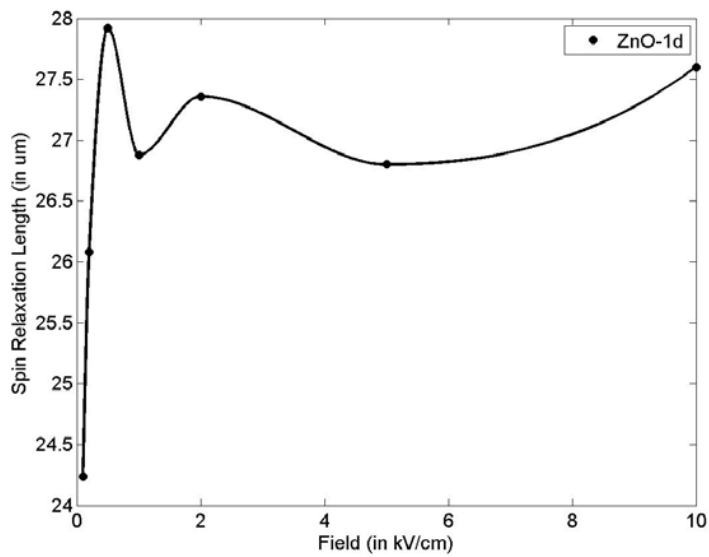 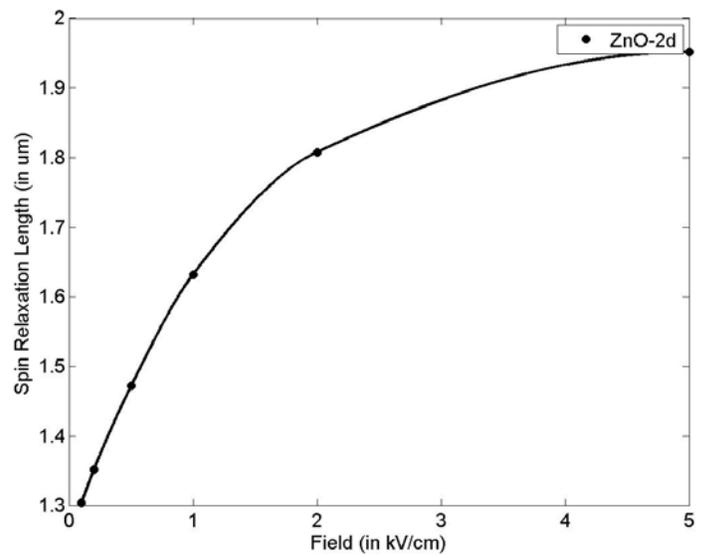

(a) (b)

Fig.14. Variation of Spin relaxation length with driving electric field for ZnO (a) 1-D nanowire (b) 2-D channel for injected electrons polarized along the z-direction at T = 300K

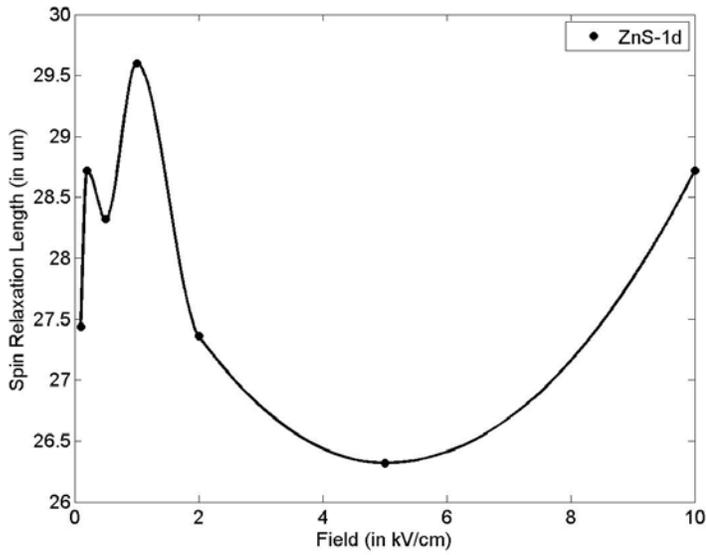 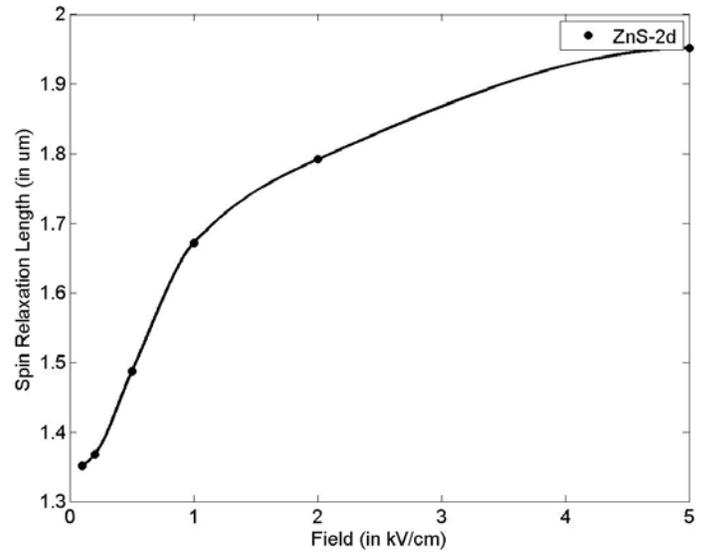

(a)                                                      (b)

Fig.15. Variation of Spin relaxation length with driving electric field for ZnS (a) 1-D nanowire (b) 2-D channel for injected electrons polarized along the z-direction at T = 300K

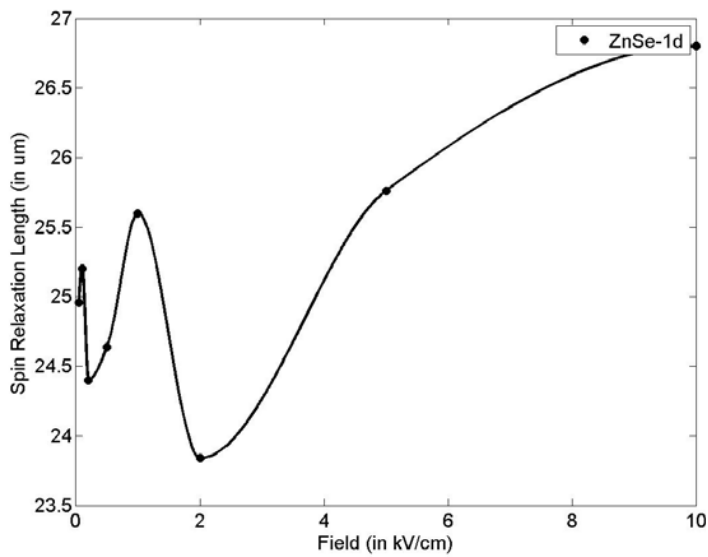 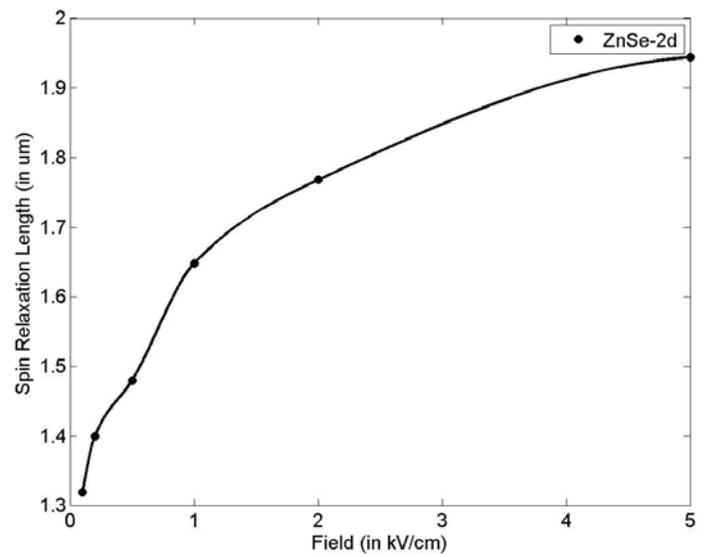

(a)                                                     (b)

Fig.16. Variation of Spin relaxation length with driving electric field for ZnSe (a) 1-D nanowire (b) 2-D channel for injected electrons polarized along the z-direction at T = 300K

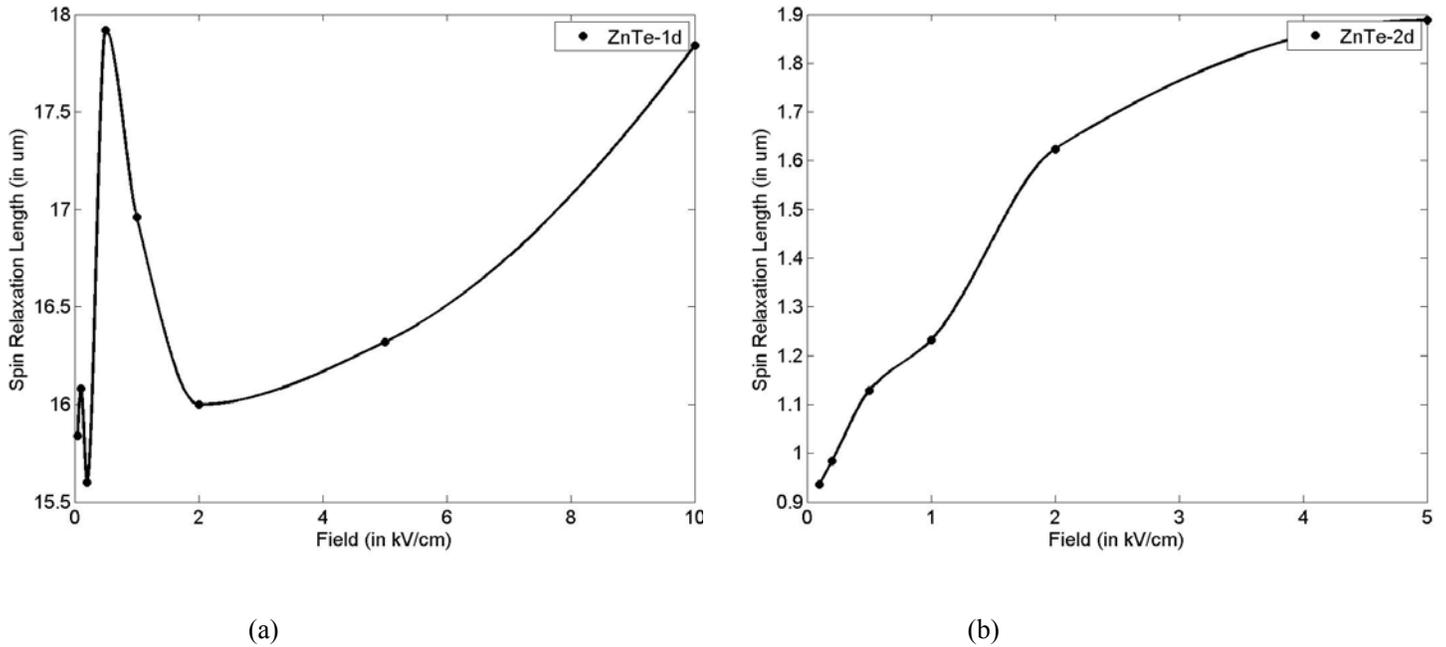

Fig.17. Variation of Spin relaxation length with driving electric field for ZnTe (a) 1-D nanowire (b) 2-D channel for injected electrons polarized along the z-direction at T = 300K

## Conclusion

We have investigated spin relaxation properties of II-VI semiconductor materials for nanowire structures and 2-D channels. We compared these types of channels and found that spin relaxation lengths in nanowires is longer by orders of magnitude. Among the various materials that we compared, ZnS showed promise as the best medium for spin transport showing the longest spin relaxation length. Variation with field followed a general trend differing for Zn compounds and Cd compounds only at high field values. These behaviour patterns also let us choose suitable operating regions or to choose suitable compound depending on strength of field being used.